\documentclass[11pt]{article}

\usepackage{amssymb,epsfig,fullpage}

\usepackage{subfigure}
\usepackage{graphicx}

\begin{document}

\title{The role of tunable activation thresholds in the dynamics of autoimmunity}

\author{K.B. Blyuss\thanks{Corresponding author. Email: k.blyuss@sussex.ac.uk}\\\\
Department of Mathematics, University of Sussex,\\Falmer, Brighton, BN1 9QH, United Kingdom\\\\
L.B. Nicholson\\\\
School of Cellular and Molecular Medicine \& School of Clinical Sciences,\\
University of Bristol, University Walk, Bristol, BS8 1TD, UK}

\maketitle

\begin{abstract}
It has been known for some time that human autoimmune diseases can be triggered by viral infections. Several possible mechanisms of interactions between a virus and immune system have been analysed, with a prevailing opinion being that the onset of autoimmunity can in many cases be attributed to "molecular mimicry", where linear peptide epitopes, processed from viral proteins, mimic normal host self proteins, thus leading to a cross-reaction of immune response against virus with host cells. In this paper we present a mathematical model for the dynamics of an immune response to a viral infection and autoimmunity, which takes into account T cells with different activation thresholds. We show how the infection can be cleared by the immune system, as well as how it can lead to a chronic infection or recurrent infection with relapses and remissions. Numerical simulations of the model are performed to illustrate various dynamical regimes, as well as to analyse the potential impact of treatment of autoimmune disease in the chronic and recurrent states. The results provide good qualitative agreement with available data on immune responses to viral infections and progression of autoimmune diseases.

\end{abstract}

\section{Introduction}

A successful immune system relies on its ability to discriminate between cells infected with a pathogen such as virus and the uninfected cells of the host. The breakdown in discrimination of self-antigens results in autoimmunity, where the immune system attacks specific cells or organs. The specificity of the immune response focuses disease within different target organs, for example, pancreatic $\beta$-cells in insulin-dependent diabetes mellitus type-1 (IDDM), the central nervous system in multiple sclerosis (MS), or the retina in uveitis - an inflammation of the eye \cite{SM,Prat,Kerr}.

Multiple factors are known to contribute toward the onset and development of autoimmune diseases, including genetic predisposition, age, and environment. Amongst the environmental factors, the major identified triggers of autoimmunity are believed to be infectious pathogens \cite{Blue,Bul}. Experimentally, direct infection of islet cells in the pancreas led to bystander damage of the islet cells and autoimmunity, and it was concluded that the release of sequestered antigen was critical to the development of disease \cite{H98}. There is also a strong association between infection with hepatitis C and autoimmune hepatitis \cite{L10}. While it is certainly not the case that {\it all} autoimmune disease is triggered by infection of the organ that subsequently becomes the target, this is a reasonable model for several known examples of organ specific autoimmunity.

There are many mechanisms by which host infection by a pathogen may contribute to autoimmunity, including triggering the innate immune system, molecular mimicry and bystander activation (see, e.g., \cite{Erc} for a recent review). Molecular mimicry is thought to be particularly important when viruses cause autoimmunity \cite{Erc, Fuji, Mu, vHO}. An immune response elicited against a viral protein that mimics a self antigen will not only eliminate the virus, but can also target normal host cells that display the cross-reactive self-antigen. Such interactions may play a role both in causing disease and also in precipitating a relapse of disease. Between such triggering events, disease can of remit in part because of the existence of regulatory pathways that dampen immune responses.

Whilst significant advances have been made in mathematical modelling of various aspects of general virus dynamics and the interactions between viruses and the immune system of the host, theoretical studies of autoimmunity have been quite limited in comparison. One of the earliest mathematical models of autoimmunity did not explicitly include specific causes of autoimmunity but rather concentrated on the interactions between effector and regulator cells, and used this to get an insight into T cell vaccination \cite{Seg}. Further progress in the context of T cell vaccination was made by Borghans et al. \cite{Bor1, Bor2}, who demonstrated how the interactions of autoreactive and regulatory T cells can lead to the onset of autoimmunity or stable oscillations around a vaccinated state. Burroughs et al. \cite{Bur1} have studied the onset of autoimmunity through bystander activation.
Wodarz and Jansen \cite{WJ} analysed autoimmunity in the context of viral causes of cancer. They included viral infections indirectly through an increased rate of uptake of self-antigen by antigen-presenting cells. Le\'on et al. \cite{Le1,Le2,Le3} have studied the dynamics of interactions between different T cells for the purposes of regulation of immune response and control of autoimmune reaction. More recently, Iwami et al. \cite{Iw1,Iw2} derived and studied a model for autoimmunity, which makes explicit account of the virus dynamics and its interaction with the immune system by means of linear or nonlinear immune response. Despite its simplicity, this model appears unable to reproduce a normal clearance of virus during a single infection, as it does not allow for a viral expansion. Various roles played by the regulatory T cells in the dynamics of autoimmunity have recently been analysed by Alexander \& Wahl \cite{AW}.

There are several ways to account for the ability of T cells to discriminate between cells presenting self antigens and infected cells. One of these is through regulatory cells, which are triggered by autoantigens and inhibit the activity of autoreactive T cells. This approach has already been extensively used in models of immune response, see Alexander \& Wahl \cite{AW}, Burroughs et al. \cite{Bur1} etc. Another approach is to consider T cells, which can perform a wide range of immune function by virtue of having different or tunable activation thresholds (TAT). This concept has been defined previously as: "activation is a threshold phenomenon and the threshold is tuned by the stimulatory experience of the cell" \cite{GP}, i.e. T cells continually tune their responsiveness to T cell antigen receptor (TCR) stimulation through stimuli evoked by autoantigens. Because the degree of autoreactivity of T cells is continuously controlled through their activation and tuning, this approach provides another way of modelling the onset and development of autoimmune disease. Grossman and Paul \cite{GP,GP2}, and Grossman and Singer \cite{GS} developed models with tunable activation thresholds that were applied to peripheral and to central T cell activation. Altan-Bonnet and Germain have modelled signalling threshold and shown differences in activation/response threshold that are dependent on the activation state of the T cell \cite{AG}. Noest \cite{NO00} has shown how the need for activation threshold tuning arises from the first principles of signal detection
theory, see also Scherer et al. \cite{SNB04} for further discussion of this issue. van den Berg and Rand have studied mathematically two cellular response models of the dynamics of tunable activation threshold \cite{VdB04}. Carneiro et al. \cite{Car05} have performed a comparative study of two mechanisms of self-tolerance: tuning of activation thresholds and control by specific regulatory T cells. The authors have shown that these two mechanisms are complementary and together provide a plausible explanation of the observed dynamics of immune tolerance. Besides purely theoretical studies,  dynamical changes in T cell activation during their circulation have also been shown experimentally both in the mouse and in man, where it has important implications for the outcome of specific therapeutic interventions \cite{Bit02,NAK00, Rom, SDG02}.

In this paper we propose and study a mathematical model for autoimmune disease caused by viral infections through molecular mimicry. By introducing separate populations of regular activated T-cells and T-cells with a lower activation threshold to self-antigen which arises as a result of the infection, the model is able to qualitatively reproduce normal aspects of immune behaviour. The model presented in this paper differs from an earlier work in this particular modelling aspect, as well as in the explicit account for a viral infection represented both by infected cells, and by a separate population of free virions.

The organization of this paper is as follows. In the next section we discuss various biological assumptions behind the origins of autoimmunity and derive the corresponding mathematical model. In Section 3 we perform a systematic analysis of the steady states of the model and their stability. Section 4 contains results of numerical simulations in different parameter regimes, which illustrate clearance of infection, onset of autoimmunity, periodic flare-ups, as well as the dynamics of multiple infections. The paper concludes with the discussion of obtained results in Section 5.

\section{The model}

We are interested in modelling the interactions between a viral infection and a human host immune system, that lead to the onset of autoimmunity through molecular mimicry. When cells in a particular organ of the body become infected with a pathogen, the elicited immune response may cross-react with one or more self-antigens that share determinants with  the pathogen \cite{vHO}. This subsequently can lead to a breakdown of tolerance for self antigens through the appearance of lymphocytes capable of attacking the host's own cells, both in the same and in other organs. For both viral and bacterial pathogens, there is evidence that their antigens derived from pathogens engage and expand both CD4 and CD8 T cells that drive autoimmune disease \cite{C11,OL}. Furthermore, recent genetic studies of autoimmunity implicate many genes that control T cell expansion and activation threshold \cite{Cot11}. Hence, it is important to understand a model of autoimmune disease that specifically addresses the dynamics of T cells. While it is unarguable that antibodies are important in many types of autoimmune disease, there is good evidence that T cell recognition of antigenic peptides is often a critical initiating step. For example, in a model of rheumatoid arthritis, antibodies were sufficient to induce disease \cite{Kor99} but the development of antibodies depends on prior T cell interactions with bacteria \cite{Wu}. Furthermore, in some of the experimental models of autoimmunity, B cells are dispensable, and disease develops when they are not present \cite{Wolf}. Therefore, the balance of evidence is for a necessary T cell component in the onset and development of autoimmunity, and our model specifically seeks to address this.

To model the dynamics of an immune response during a viral infection and possible onset of autoimmunity, we employ a model similar to those studied earlier in the context of immune responses \cite{NM, WJ}. Let $A$ denote the number of susceptible cells in a particular organ or tissue. We consider two distinct populations of susceptible cells, $A_1$ and $A_2$ to allow for a situation when autoimmunity takes place in a different organ to the one where the original infection occurs. It is assumed that in the absence of infection or autoimmunity, these two cell populations would be maintained at a certain constant level supported by homeostasis. It has been previously shown \cite{Iw1,Iw2} that a specific form of the growth function for susceptible cells can have a significant effect on the overall dynamics of autoimmune diseases, and here we use the logistic form, as studied in Iwami et al. \cite{Iw1}, and by Perelson and Nelson \cite{PN} in models for HIV infection.

When a person acquires a viral infection, a proportion of cells become infected with this virus, and after a certain period of time the infected cells, whose population is denoted by $F$, will start producing virions (free virus particles) $V$ that will go on to infect other as yet uninfected cells $A_2$ (and, possibly, $A_1$). All time constants, such as the time to encounter an uninfected cell and a time required for cell entry, are implicitly included in the rate of infection $\lambda$. In the absence of infection, the na\"ive T-cells $T_{in}$ are taken to follow a logistic growth.  Once activated, $T_{in}$ cells become $T_1$ cells that have the ability to kill infected cells $F$. For a fraction of these cells $T_2$ their activation threshold for stimulation by susceptible cells that are not infected is reduced, allowing them to kill $A$ cells. This is the autoimmune response. It has already been shown that T cells with varying activation threshold can have a significant effect on the dynamics of the immune response \cite{GP,GS}.

With the above assumptions, the model to be analysed in this paper has the form
\begin{equation}\label{model}
\begin{array}{l}
\displaystyle{\frac{dA_1}{dt}=r_1 A_1\left(1-\frac{A_1}{N_1}\right)-p_1 \lambda A_1V-\alpha_{a}T_{2}A_1,}\\\\
\displaystyle{\frac{dA_2}{dt}=r_2 A_2\left(1-\frac{A_2}{N_2}\right)-\lambda A_2 V-p_2 \alpha_{a}T_{2}A_2,}\\\\
\displaystyle{\frac{dF}{dt}=\lambda (p_1A_1+A_2)V-\mu_{F}F-\alpha_{F}T_{1}F-\alpha_{a}T_{2}F,}\\\\
\displaystyle{\frac{dT_{in}}{dt}=g_{t}T_{in}\left(1-\frac{T_{in}}{M}\right)-\alpha_{act}T_{in}F,}\\\\
\displaystyle{\frac{dT_{1}}{dt}=\alpha_{act}T_{in}F-T_{1}(\mu_{1}+\tau),}\\\\
\displaystyle{\frac{dT_{2}}{dt}=\tau T_{1}-\mu_{2}T_{2},}\\\\
\displaystyle{\frac{dV}{dt}=kF-\gamma V,}
\end{array}
\end{equation}
where $A_{1,2}$, $F$, $T_{in}$, $T_1$, $T_2$ and $V$ denote the populations of susceptible cells (possibly, affected by infection and autoimmunity separately), infected cells, na\"ive T cells, activated T cells, T cells with a lower activation threshold to self-antigen, and free virus, respectively. As it has already been mentioned, in the absence of infection which can trigger autoimmune reaction, the susceptible cells $A_{1,2}$ reproduce logistically with their respective linear growth rates $r_{1,2}$ and carrying capacities $N_{1,2}$. These cells become infected at a rate $\lambda$, and they are destroyed at a rate $\alpha_{a}$ by autoreactive T-cells $T_{2}$ with a lower activation threshold to self-antigen (this is the actual implementation of the autoimmunity mechanism). Na\"ive T-cells $T_{in}$ are assumed to have a logistic growth with a linear growth rate $g_{t}$ and a carrying capacity $M$; they get primed by dendritic cells at a rate taken to be proportional to the number of infected cells with a constant $\alpha_{act}$.

Once infected, cells presenting foreign antigen $F$ die at a rate $\mu_{F}$, and they are also destroyed by the activated T-cells $T_{1}$ at a rate $\alpha_{F}$, and by the autoreactive T-cells $T_{2}$ at a rate $\alpha_{a}$.  Activated T-cells $T_{1}$ die at a rate $\mu_{1}$, and at a rate $\tau$ they produce autoreactive T cells $T_{2}$ with a lower activation threshold to self-antigen. Finally, the T cells $T_{2}$ die at a rate $\mu_{2}$. Free virions $V$ are  produced by an infected cell at a rate $k$, and $\gamma$  is the natural clearance rate of the virus.

Parameters $0\leq p_1,p_2\leq 1$ control whether the populations of susceptible cells are affected by autoimmune reaction and/or infection. When $p_1=p_2=0$, only cells $A_2$ are affected by the infection, and the autoimmunity only affects the population $A_1$ of other cells. Biologically, this situation arises when a pathogen causes infection in one part of the body, and the autoimmune response causes damage in other parts of the body. If $p_1=0$ and $p_2>0$, only the cell population $A_2$ experiences infection, but both of the susceptible cell populations are affected by the developing autoimmunity. Conversely, if $p_2=0$ and $p_1>0$, both $A_1$ and $A_2$ cells become infected, and only $A_1$ also experiences autoimmunity. Finally, in the case of $p_{1,2}>0$, both of the cell populations $A_1$ and $A_2$ get exposed to both the infection and the autoimmune reaction, although they can potentially be affected by infection/autoimmunity at different rates, and the differences in intrinsic growth rates and carrying capacities of these two cell populations can also lead to a significantly different dynamics of these two cell populations.

The main emphasis of the above model is on the separation of T cell populations into two activated populations, one of which is capable of an autoimmune reaction through having a lower activation threshold to self-antigen. As we are particularly interested in the role of foreign infections in the onset of autoimmunity, we have not explicitly included in the model several other aspects that can be of interest in specific contexts, such as antibody response or regulatory T cells. Furthermore, we have not included memory cells in our model, which can be important in the analysis of a longer-term dynamics or multiple infections.

\section{Steady states}

We begin our analysis of the system (\ref{model}) by considering its steady states $E^*=(A_1^*,A_2^*,F^*,T_{in}^*,T_1^*,T_2^*,V^*)$, which can be found by equating the right-hand side of system (\ref{model}) to zero. Stability is determined by the eigenvalues of the Jacobian of linearization of the system (\ref{model}) near each of the steady states. In order to systematically study the steady states of system (1), we first consider all steady states with $V^*=0$ (this immediately implies $F^*=T_1^*=T_2^*=0$). There are exactly eight of such steady states:
\begin{equation}
\begin{array}{ll}
E_1^*=(0,0,0,0,0,0,0),& \hspace{1cm}E_5^*=(N_1,0,0,0,0,0,0),\\\\
E_2^*=(0,0,0,M,0,0,0),& \hspace{1cm}E_6^*=(N_1,0,0,M,0,0,0),\\\\
E_3^*=(0,N_2,0,0,0,0,0),& \hspace{1cm}E_7^*=(N_1,N_2,0,0,0,0,0),\\\\
E_4^*=(0,N_2,0,M,0,0,0),& \hspace{1cm}E_8^*=(N_1,N_2,0,M,0,0,0).
\end{array}
\end{equation}
Computing the Jacobian at each of these steady states shows that $E_1^*$ to $E_7^*$ are always saddles (and hence unstable) for any values of the system parameters. The steady state $E_8^*$ is a stable node, provided
\[
k\lambda(N_2+p_1N_1)<\mu_F\gamma,
\]
or a saddle if
\[
k\lambda(N_2+p_1N_1)>\mu_F\gamma.
\]
At
\[
k\lambda(N_2+p_1N_1)=\mu_F\gamma,
\]
the steady state $E_8^*$ undergoes a steady state bifurcation, where one of the eigenvalues goes through zero along the real axis. Biologically, the above stability condition represents the rate of change in the number of infected cells being negative at the steady state.

When $V^*\neq 0$, we can have either $A_1^*=0$ and $A_2^*\neq 0$, or $A_1^*\neq 0$ and $A_2^*=0$, or $A_1^*\neq 0$ and $A_2\neq 0$. In the case where $A_1^*=0$ and $A_2^*\neq 0$, there are two options. One of these is $E_9^*$ with $T_{in}^*=T_1^*=T_2^*=0$, in which case we also have
\[
A_2^*=\frac{\mu_F\gamma}{\lambda k},\hspace{0.5cm}F^*=\frac{r_2\gamma}{\lambda k}\left(1-\frac{\mu_F\gamma}{\lambda kN_2}\right),
\hspace{0.5cm}V^*=\frac{r_2}{\lambda}\left(1-\frac{\mu_F\gamma}{\lambda kN_2}\right).
\]
The steady state $E_9^*$ with such values of the variables is stable, provided the following conditions are satisfied
\begin{equation}
\begin{array}{l}
p_1r_2N_2\lambda k>\lambda kr_1N_2+p_1r_2\gamma\mu_F,\hspace{0.3cm}\alpha_{act}r_2\gamma \lambda kN_2>g_t\lambda^2k^2N_2+\alpha_{act}r_2\gamma^2\mu_F,\\\\
\gamma\mu_Fr_2(\mu_F+\gamma)+\lambda kN_2[(\mu_F+\gamma)^2+\mu_F\gamma-\lambda kN_2]>0.
\end{array}
\end{equation}
It is noteworthy that when $p_1=0$, this steady state is unstable for {\it any values} of other parameters. For $p_1=1$, this steady state can only be stable, provided the linear growth rate $r_2$ sufficiently exceeds $r_1$. The steady state $E_9^*$ can undergo a steady state bifurcation when
\begin{equation}\label{E9ss}
p_1r_2N_2\lambda k=\lambda kr_1N_2+p_1r_2\gamma\mu_F,\hspace{0.3cm}\mbox{ or }\hspace{0.3cm}\alpha_{act}r_2\gamma \lambda kN_2=g_t\lambda^2k^2N_2+\alpha_{act}r_2\gamma^2\mu_F,
\end{equation}
and it can also undergo a Hopf bifurcation, when the system parameters satisfy the conditions
\begin{equation}\label{E9Hopf}
\begin{array}{l}
p_1r_2N_2\lambda k>\lambda kr_1N_2+p_1r_2\gamma\mu_F,\hspace{0.3cm}\alpha_{act}r_2\gamma \lambda kN_2>g_t\lambda^2k^2N_2+\alpha_{act}r_2\gamma^2\mu_F,\\\\
\gamma\mu_Fr_2(\mu_F+\gamma)+\lambda kN_2[(\mu_F+\gamma)^2+\mu_F\gamma-\lambda kN_2]=0.
\end{array}
\end{equation}
An important observation is that the manifold $A_1=T_{in}=T_1=T_2=0$ is flow-invariant, and the Hopf bifurcation of the steady state $E_9^*$ takes place inside this manifold, which results in the appearance of a periodic orbit confined to the same manifold. The importance of this observation lies in the fact that this periodic solution does not cause oscillations of any of the variables which would go below zero, making it unrealistic.\\

The second option for the case $A_1^*=0$ and $A_2^*\neq 0$ is given by the steady state $E_{10}^*$, for which $T_{in}^*$ satisfies the quadratic equation
\begin{equation}\label{Quad1}
\begin{array}{l}
g_t\gamma\alpha_{act}[r_2\gamma(\alpha_F\mu_2+\alpha_a\tau)+p_2\alpha_a\tau N_2\lambda k](T_{in}^*)^2\\\\
-g_t[r_2\gamma^2
\alpha_{act}M(\alpha_F\mu_2+\alpha_a\tau)-\lambda^2k^2N_2\mu_2(\mu_1+\tau)+p_2\lambda\tau\gamma\alpha_a\alpha_{act} kMN_2]T_{in}^*\\\\
+\mu_2M(\mu_1+\tau)[r_2\gamma\alpha_{act}(N_2\lambda k-\mu_F\gamma)-\lambda^2k^2g_tN_2]=0,
\end{array}
\end{equation}
and the other variables are given by
\[
\begin{array}{l}
\displaystyle{A_2^*=\frac{\gamma}{\lambda k}\left[\mu_F+\frac{\alpha_F\mu_2+\alpha_a\tau}{\mu_2(\mu_1+\tau)}g_tT_{in}^*\left(1-\frac{T_{in}^*}{M}\right)\right],}\\\\
\displaystyle{F^*=\frac{g_t}{\alpha_{act}}\left(1-\frac{T_{in}^*}{M}\right),\hspace{0.5cm}V^*=\frac{kg_t}{\gamma\alpha_{act}}\left(1-\frac{T_{in}^*}{M}\right),}\\\\
\displaystyle{T_1^*=\frac{g_t}{\mu_1+\tau}T_{in}^*\left(1-\frac{T_{in}^*}{M}\right),\hspace{0.5cm}T_2^*=\frac{\tau g_t}{\mu_2(\mu_1+\tau)}T_{in}^*\left(1-\frac{T_{in}^*}{M}\right).}
\end{array}
\]
Depending on the particular values of parameters there can be between zero and two distinct steady states determined by the roots of equation (\ref{Quad1}). It does not prove possible to determine stability of the steady state $E_{10}^*$ in a closed form, and hence one has to compute the eigenvalues of the corresponding Jacobian numerically.\\

In a similar way, when $A_1^*\neq 0$ and $A_2^*=0$, there are again two options. The first one, denoted by $E_{11}^*$, describes the case when $T_{in}^*=T_1^*=T_2^*=0$, and the other variables can be found as
\[
A_1^*=\frac{\mu_F\gamma}{\lambda p_1k},\hspace{0.5cm}F^*=\frac{r_1\gamma}{\lambda p_1k}\left(1-\frac{\mu_F\gamma}{\lambda p_1kN_1}\right),
\hspace{0.5cm}V^*=\frac{r_1}{\lambda p_1}\left(1-\frac{\mu_F\gamma}{\lambda p_1kN_1}\right).
\]
The steady state $E_{11}^*$ is stable, provided the following conditions are satisfied
\begin{equation}
\begin{array}{l}
p_1\lambda kr_1N_1>p_1^2\lambda kr_2N_1+r_1\gamma\mu_F,\alpha_{act}p_1r_1\gamma \lambda kN_1>p_1^2g_t\lambda^2k^2N_1+\alpha_{act}r_1\gamma^2\mu_F,\\\\
\gamma\mu_Fr_1(\mu_F+\gamma)+p_1\lambda kN_1[(\mu_F+\gamma)^2+\mu_F\gamma-p_1\lambda kN_1]>0.
\end{array}
\end{equation}
The steady state $E_{11}^*$ can undergo a steady state bifurcation when
\begin{equation}\label{E11ss}
p_1\lambda kr_1N_1=p_1^2\lambda kr_2N_1+r_1\gamma\mu_F,\mbox{ or }\alpha_{act}p_1r_1\gamma \lambda kN_1=p_1^2g_t\lambda^2k^2N_1+\alpha_{act}r_1\gamma^2\mu_F,
\end{equation}
and it can also undergo a Hopf bifurcation, when the system parameters satisfy the conditions
\begin{equation}\label{E11Hopf}
\begin{array}{l}
p_1\lambda kr_1N_1>p_1^2\lambda kr_2N_1+r_1\gamma\mu_F,\alpha_{act}p_1r_1\gamma \lambda kN_1>p_1^2g_t\lambda^2k^2N_1+\alpha_{act}r_1\gamma^2\mu_F,\\\\
\gamma\mu_Fr_1(\mu_F+\gamma)+p_1\lambda kN_1[(\mu_F+\gamma)^2+\mu_F\gamma-p_1\lambda kN_1]=0.
\end{array}
\end{equation}
Similarly to the case of $E_9^*$, the Hopf bifurcation of the steady state $E_{11}^*$ takes place inside the flow-invariant manifold $A_2=T_{in}=T_1=T_2=0$. Analogously to the analysis of stability of the steady state $E_{9}$, one can note that for stability of the steady state $E_{11}$, the linear growth rate $r_1$ should be sufficiently greater than the rate $r_2$.

The second option for the case $A_1^*\neq 0$ and $A_2^*=0$ is $E_{12}^*$, for which $T_{in}^*$ is different from zero and satisfies the quadratic equation
\begin{equation}\label{Quad2}
\begin{array}{l}
g_t\gamma\alpha_{act}[r_1\gamma(\alpha_F\mu_2+\alpha_a\tau)+p_1\alpha_a\tau N_1\lambda k](T_{in}^*)^2\\\\
-g_t[r_1\gamma^2
\alpha_{act}M(\alpha_F\mu_2+\alpha_a\tau)-p_1^2\lambda^2k^2N_1\mu_2(\mu_1+\tau)+p_1\lambda\tau\gamma\alpha_a\alpha_{act} kMN_1]T_{in}^*\\\\
+\mu_2M(\mu_1+\tau)[r_1\gamma\alpha_{act}(p_1N_1\lambda k-\mu_F\gamma)-p_1\lambda^2k^2g_tN_1]=0,
\end{array}
\end{equation}
with the other variables being given by
\[
\begin{array}{l}
\displaystyle{A_1^*=\frac{\gamma}{p_1\lambda k}\left[\mu_F+\frac{\alpha_F\mu_2+\alpha_a\tau}{\mu_2(\mu_1+\tau)}g_tT_{in}^*\left(1-\frac{T_{in}^*}{M}\right)\right],}\\\\
\displaystyle{F^*=\frac{g_t}{\alpha_{act}}\left(1-\frac{T_{in}^*}{M}\right),\hspace{0.5cm}V^*=\frac{kg_t}{\gamma\alpha_{act}}\left(1-\frac{T_{in}^*}{M}\right),}\\\\
\displaystyle{T_1^*=\frac{g_t}{\mu_1+\tau}T_{in}^*\left(1-\frac{T_{in}^*}{M}\right),\hspace{0.5cm}T_2^*=\frac{\tau g_t}{\mu_2(\mu_1+\tau)}T_{in}^*\left(1-\frac{T_{in}^*}{M}\right).}
\end{array}
\]

Finally, when both $A_1^*\neq 0$ and $A_2^*\neq 0$, there are again two options. The first one corresponds to a steady state $E_{13}^*$ with $T_{in}^*=T_1^*=T_2^*=0$, and the other variables being given by
\[
\begin{array}{l}
\displaystyle{F^*=\frac{r_1r_2\gamma\left[(p_1N_1+N_2)\lambda k-\mu_F \gamma\right]}{\lambda^2 k^2(r_1N_2+p_1^2r_2N_1)},\hspace{0.3cm}
V^*=\frac{r_1r_2 \left[(p_1N_1+N_2)\lambda k-\mu_F \gamma\right]}{\lambda^2 k(r_1N_2+p_1^2r_2N_1)},}\\\\
\displaystyle{A_1^*=\frac{N_1[\lambda kN_2(r_1-p_1r_2)+\mu_F\gamma r_2p_1]}{\lambda k(r_1N_2+p_1^2r_2N_1)},
A_2^*=\frac{N_2[\lambda kp_1N_1(p_1r_2-r_1)+\mu_F\gamma r_1]}{\lambda k(r_1N_2+p_1^2r_2N_1)}.}
\end{array}
\]
Numerical computation of the eigenvalues of linearization shows that for sufficiently small values of $\gamma$, the steady state $E_{13}^*$ is stable; as $\gamma$ increases, stability is lost, but then regained again as $\gamma$ increases further still.

The final possibility corresponds to a steady state $E_{14}^*$ with the values of all variables being different from zero, so that $T_{in}^*$ satisfies the quadratic equation
\begin{equation}
\begin{array}{l}
g_t\gamma\alpha_{act}\left[\lambda\alpha_a\tau k(p_1r_2N_1+p_2r_1N_2)+\gamma r_1r_2(\alpha_F\mu_2+\alpha_a\tau)\right](T_{in}^*)^2\\\\
-g_t\Big\{M\gamma\alpha_{act} [\lambda k\alpha_a\tau(p_1r_2N_1+p_2r_1N_2)+\gamma r_1r_2(\alpha_F\mu_2+\alpha_a\tau)]\\\\
-\lambda^2k^2\mu_2(\mu_1+\tau)(p_1^2r_2N_1+r_1N_2)\Big\}T_{in}^*-\mu_2M(\mu_1+\tau)[\lambda^2 k^2g_t(p_1^2r_2N_1+r_1N_2)\\\\
-\lambda k\gamma\alpha_{act}r_1r_2(p_1N_1+N_2)+\mu_F\gamma^2\alpha_{act}r_1r_2]=0,
\end{array}
\end{equation}
and the other variables can be found as
\[
\begin{array}{l}
\displaystyle{A_1^*=\frac{N_1\left[\mu_2M(\mu_1+\tau)(r_1\gamma\alpha_{act}-p_1\lambda kg_t)+\gamma\alpha_a\alpha_{act}\tau g_tT_{in}(T_{in}^*-M)\right]}{M\mu_2\gamma\alpha_{act}r_1(\mu_1+\tau)},}\\\\
\displaystyle{A_2^*=\frac{N_2\left[\mu_2(\mu_1+\tau)(r_2\gamma\alpha_{act}M+\lambda kg_t(T_{in}^*-M))+p_2\gamma\alpha_a\alpha_{act}\tau g_tT_{in}^*(T_{in}^*-M)\right]}{M\mu_2\gamma\alpha_{act}r_2(\mu_1+\tau)},}\\\\
\displaystyle{F^*=\frac{g_t}{\alpha_{act}}\left(1-\frac{T_{in}^*}{M}\right),\hspace{0.5cm}V^*=\frac{kg_t}{\gamma\alpha_{act}}\left(1-\frac{T_{in}^*}{M}\right),}\\\\
\displaystyle{T_1^*=\frac{g_t}{\mu_1+\tau}T_{in}^*\left(1-\frac{T_{in}^*}{M}\right),\hspace{0.5cm}T_2^*=\frac{\tau g_t}{\mu_2(\mu_1+\tau)}T_{in}^*\left(1-\frac{T_{in}^*}{M}\right).}
\end{array}
\]
\begin{figure}
\hspace{-1cm}
\epsfig{file=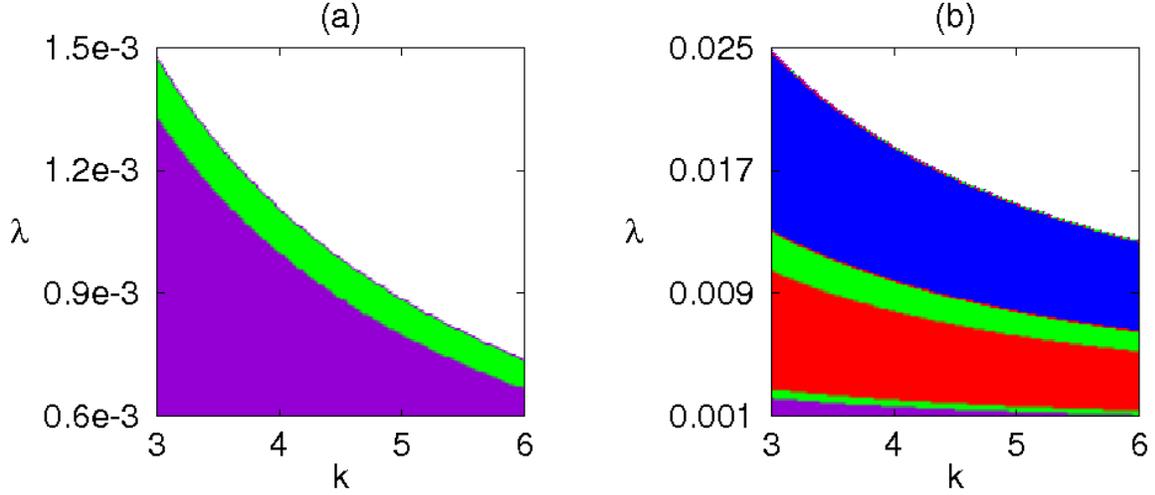,width=18cm}
\caption{Stability regions of different steady states depending on $\lambda$ and $k$. Parameter values are $r_1=0.2$, $r_2=0.1$, $N_1=200$, $N_2=300$, $M=300$,  $g_t=0.2$, $\alpha_{F}=0.0006$, $\alpha_{act}=0.04$, $\alpha_a=0.033$, $\mu_F=1$, $\mu_{1}=0.08$, $\mu_{2}=0.02$, $\tau=0.001$. (a) $p_1=p_2=1$. (b) $p_1=p_2=0$.
Colour code denotes region of stability of a single steady state: $E_{8}^*$ (violet), $E_{14}^*$ (green), $E_{13}^*$ (red),  
$E_{10}^*$ (blue), and white denotes a parameter region where all steady states are unstable.}\label{stab_ss}
\end{figure}

To illustrate how stability of different steady states is affected by the system parameters, we show in Fig.~\ref{stab_ss} regions in the $k$-$\lambda$ parameter plane where different steady states are stable depending on whether infection and autoimmunity affect the same of different cell populations. This figure suggests that for sufficiently small rate of infection $\lambda$, the only possible stable steady state is  the disease-free state $E_8^*$, in which cell populations $A_1$, $A_2$ and $T_{in}$ are maintained at their respective carrying capacities, and there no free virus or infection, thus implying the absence of activated T cells $T_1$ and $T_2$. As the infection rate $\lambda$ increases, this disease-free steady state loses stability, and the system turns to a stable state of chronic infection $E_{14}$, where all cell populations are different from zero. When this steady states loses stability in the case of $p_1=p_2=1$ (i.e. when both cell populations $A_1$ and $A_2$ are the targets of both infection and autoimmunity), there appear to be no other stable steady state for any values of $\lambda$ and $k$, as illustrated in plot \ref{stab_ss}(a). At the same time, it is worth noting that  some other stable steady states can be recovered when other parameters are changed. For instance, stability of the steady states $E_{9}$ and $E_{11}$ depends heavily on the relation between the two linear growth rates $r_1$ and $r_2$, so varying these may produce additional stable steady states. In the case $p_1=p_2=0$, the range of possibilities is much wider, as shown in plot \ref{stab_ss}(b). It is possible to observe stable state $E_{10}$, where the cell population $A_1$ is equal to zero with all other populations being positive. Another possibility is a state $E_{13}$, in which infection renders virus specific T cells ineffective. An important observation is that in all parameter regimes there appears to be a single stable steady state, hence no bi-stability is possible in the system. Whilst Figure~\ref{stab_ss} may not illustrate the complete catalogue of possible steady states, it provides an insight into how stability is affected by the system parameters.

We summarize the above analysis of possible steady states of the system (\ref{model}) in the following table.\\

\noindent 
\hspace{1cm}
\begin{tabular}{| l | l | l |p{11cm}}
\hline
\begin{tabular}{l}
Disease-free\\
steady states\\
$V=F=0$,\\
$T_{in}=T_1=T_2=0$
\end{tabular}& $E_{1}$ & \begin{tabular}{l}all cell populations are equal to zero,\\ always
{\it unstable} \end{tabular}\\ \cline{2-3}
& $E_2$& \begin{tabular}{l}$T_{in}=M$, all other populations are\\ equal to zero, always
{\it unstable} \end{tabular}\\ \cline{2-3}
& $E_3$& \begin{tabular}{l}$A_2=N_2$, all other populations are\\ equal to zero, always
{\it unstable} \end{tabular}\\ \cline{2-3}
& $E_4$& \begin{tabular}{l}$A_2=N_2$, $T_{in}=M$, all other populations\\ are equal to zero, always
{\it unstable} \end{tabular}\\ \cline{2-3}
& $E_5$& \begin{tabular}{l}$A_1=N_1$, all other populations are\\ equal to zero, always
{\it unstable} \end{tabular}\\ \cline{2-3}
& $E_6$& \begin{tabular}{l}$A_1=N_1$, $T_{in}=M$, all other populations\\ are equal to zero, always
{\it unstable} \end{tabular}\\ \cline{2-3}
& $E_7$& \begin{tabular}{l}$A_1=N_1$, $A_2=N_2$, all other populations\\ are equal to zero, always
{\it unstable} \end{tabular}\\ \cline{2-3}
& $E_8$& \begin{tabular}{l}$A_1=N_1$, $A_2=N_2$, $T_{in}=M$, all other\\populations are equal to zero, can be\\
{\it stable} or {\it unstable} \end{tabular}\\ \hline
\begin{tabular}{l}
Steady states\\
with a chronic\\
viral infection\\
$V>0$\end{tabular}& $E_9$ &\begin{tabular}{l} $A_1=T_{in}=T_1=T_2=0$, other\\
populations are positive, can be\\
{\it stable} or {\it unstable} \end{tabular}\\ \cline{2-3}
& $E_{10}$& \begin{tabular}{l}$A_1=0$, all other populations are\\ positive, can be
{\it stable} or {\it unstable} \end{tabular}\\ \cline{2-3}
& $E_{11}$ &\begin{tabular}{l} $A_2=T_{in}=T_1=T_2=0$, other\\
populations are positive, can be\\
{\it stable} or {\it unstable} \end{tabular}\\ \cline{2-3}
& $E_{12}$& \begin{tabular}{l}$A_2=0$, all other populations are\\ positive, can be
{\it stable} or {\it unstable} \end{tabular}\\ \cline{2-3}
& $E_{13}$ &\begin{tabular}{l} $T_{in}=T_1=T_2=0$, other populations\\ are positive, can be
{\it stable} or {\it unstable} \end{tabular}\\ \cline{2-3}
& $E_{14}$& \begin{tabular}{l}all cell populations are positive,\\ can be
{\it stable} or {\it unstable} \end{tabular}\\ \hline
\end{tabular}

\section{Numerical simulations}

In order to illustrate various dynamical regimes that can be exhibited our model, we solve the the system (\ref{model}) numerically in different parameter regimes, taking account of results concerning stability and bifurcations of the steady states analysed in the previous section. Prior to performing simulations, we simplify the system by introducing the non-dimensional variables $(\widehat{A}_1,\widehat{A}_2,\widehat{F},\widehat{T}_{in},\widehat{T}_1,\widehat{T}_2,\widehat{A}_1,\widehat{V})$ and $\widehat{t}$ as follows,
\[
\begin{array}{l}
\displaystyle{\widehat{t}=g_t t,\hspace{0.5cm}A_1=N_1\widehat{A}_1,\hspace{0.5cm}A_2=N_2\widehat{A}_2,\hspace{0.5cm}F=(N_1+N_2)\widehat{F},}\\\\
\displaystyle{T_{in}=M\widehat{T}_{in},\hspace{0.5cm}T_1=M\widehat{T}_1,\hspace{0.5cm}T_2=M\widehat{T}_2,\hspace{0.5cm}V=V_0\widehat{V}.}
\end{array}
\]
Substituting these variables into the system (\ref{model}) yields
\begin{equation}\label{nondim}
\begin{array}{l}
\displaystyle{\frac{d\widehat{A}_1}{d\widehat{t}}=\widehat{r}_1 \widehat{A}_1\left(1-\widehat{A}_1\right)-p_1 \widehat{\lambda} \widehat{A}_1\widehat{V}-\widehat{\alpha}_{a}\widehat{T}_{2}\widehat{A}_1,}\\\\
\displaystyle{\frac{d\widehat{A}_2}{dt}=\widehat{r}_2 \widehat{A}_2\left(1-\widehat{A}_2\right)-\widehat{\lambda} \widehat{A}_2 \widehat{V}-p_2 \widehat{\alpha}_{a}\widehat{T}_{2}\widehat{A}_2,}\\\\
\displaystyle{\frac{d\widehat{F}}{dt}=\widehat{\lambda}\left[p_1N_p\widehat{A}_1+(1-N_p)\widehat{A}_2\right]\widehat{V}-(\widehat{\mu}_{F}+\widehat{\alpha}_{F}\widehat{T}_{1}+\widehat{\alpha}_{a}\widehat{T}_{2})\widehat{F},}\\\\
\displaystyle{\frac{d\widehat{T}_{in}}{d\widehat{t}}=\widehat{T}_{in}\left(1-\widehat{T}_{in}\right)-\widehat{\alpha}_{act}\widehat{T}_{in}\widehat{F},}\\\\
\displaystyle{\frac{d\widehat{T}_{1}}{d\widehat{t}}=\widehat{\alpha}_{act}\widehat{T}_{in}\widehat{F}-\widehat{T}_{1}(\widehat{\mu}_{1}+\widehat{\tau}),}\\\\
\displaystyle{\frac{d\widehat{T}_{2}}{d\widehat{t}}=\widehat{\tau}\widehat{T}_{1}-\widehat{\mu}_{2}\widehat{T}_{2},}\\\\
\displaystyle{\frac{d\widehat{V}}{d\widehat{t}}=\widehat{k}\widehat{F}-\widehat{\gamma}\widehat{V},}
\end{array}
\end{equation}
where the modified parameters are given by
\[
\begin{array}{l}
\displaystyle{\widehat{r}_1=\frac{r_1}{g_t},\hspace{0.5cm}\widehat{r}_2=\frac{r_2}{g_t},\hspace{0.5cm}\widehat{\lambda}=\frac{\lambda V_0}{g_t},\hspace{0.5cm}\widehat{\gamma}=\frac{\gamma}{g_t},\hspace{0.5cm}\widehat{k}=\frac{k(N_1+N_2)}{g_t V_0},}\\\\
\displaystyle{\widehat{\alpha}_a=\frac{\alpha_a M}{g_t},\hspace{0.5cm}\widehat{\alpha}_F=\frac{\alpha_F M}{g_t},\hspace{0.5cm}\widehat{\alpha}_{act}=\frac{\alpha_{act}(N_1+N_2)}{g_t},\hspace{0.5cm}\widehat{\tau}=\frac{\tau}{g_t},}\\\\
\displaystyle{\widehat{\mu}_1=\frac{\mu_1}{g_t},\hspace{0.5cm}\widehat{\mu}_2=\frac{\mu_2}{g_t},\hspace{0.5cm}\widehat{\mu}_F=\frac{\mu_F}{g_t},}
\end{array}
\]
\begin{figure}
\hspace{-2cm}
\epsfig{file=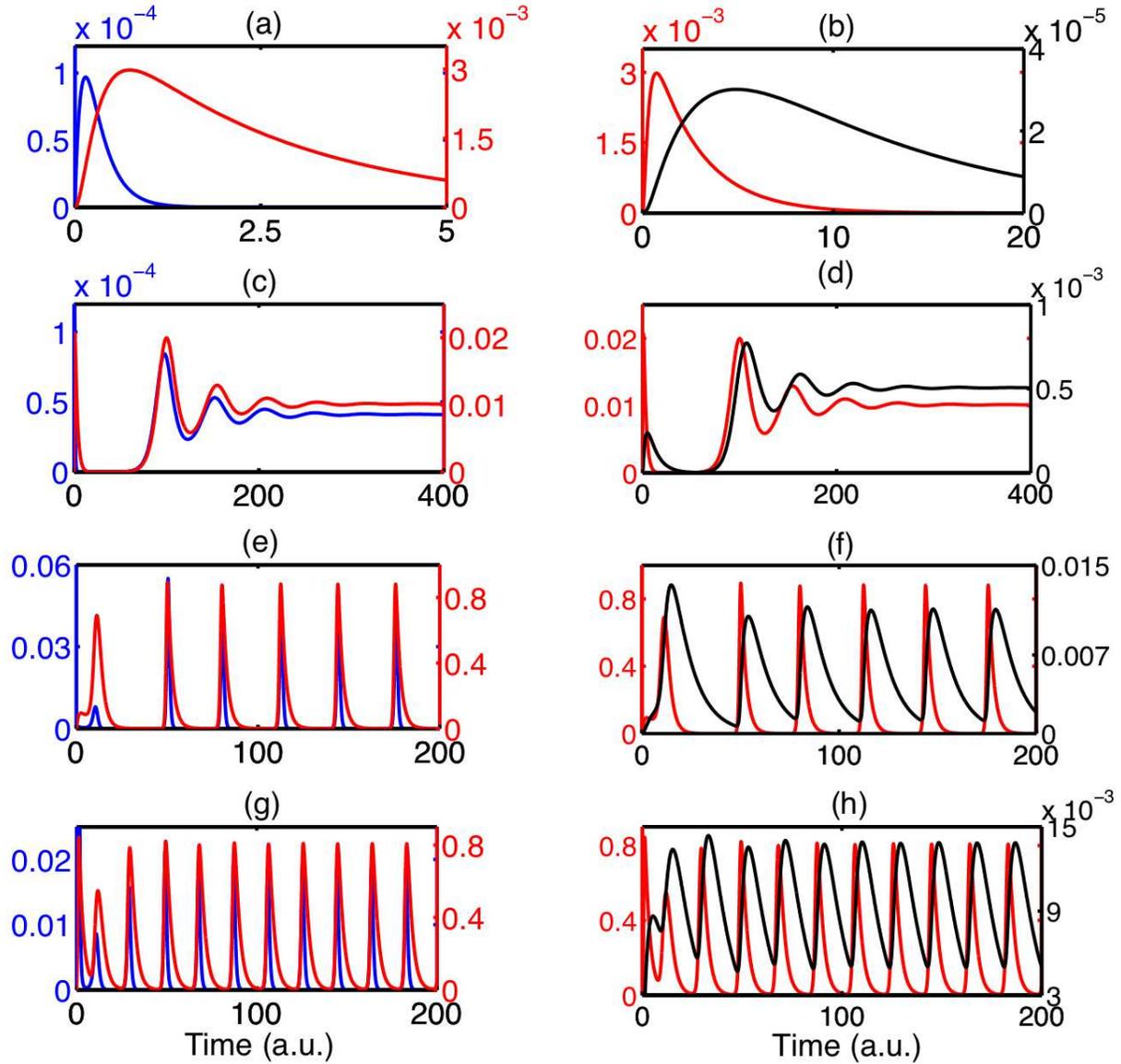,width=20cm}
\caption{Numerical solution of the system (\ref{nondim}) for $p_1=p_2=1$. Parameter values are $r_1=1$, $r_2=0.5$, 
$N_p=0.4$, $g_t=0.2$, $\alpha_a=50$, $\mu_{F}=5$, $\alpha_{act}=100$,
$\alpha_{F}=1$, $\mu_{1}=0.4$, $\mu_{2}=0.1$, $\tau=0.005$, $k=105$, $\gamma=10$. (a) and (b) $\lambda=0.1$, (c) and (d) $\lambda=0.5$, (e) and (f) $\lambda=1$, (g) and (h) $\lambda=3$. In all plots, the colours denote infected cells $F$ (blue), activated T cells $T_1$ (red), and autoreactive T cells $T_2$ (black). Time is measured in arbitrary units (a.u.).}\label{fig11}
\end{figure}
and we have introduced one additional parameter
\[
N_p=\frac{N_1}{N_1+N_2}.
\]
The above non-dimensionalization reduces the number of free parameters by three, thus reducing the overall complexity and simplifying the survey of parameter space. To simplify the notation, we will drop hats for variables and parameters in the system (\ref{nondim}). The values of system parameters used in simulations of system (\ref{nondim}) are the rescaled values of the parameters used in Fig.~\ref{stab_ss}.

\subsection{Single infection}

First, we consider the situation when the host experiences a single infection by a virus, and one is interested in the subsequent dynamics of the immune response against this infection, as well as possible autoimmune reaction. Following Wodarz {\it et al.} \cite{WJ,WL} and Vickers {\it et al.} \cite{VZO}, we define a certain threshold value, below which the infection is considered extinct; for simulations presented below this threshold was chosen to be $10^{-8}$ and applied to the number of infected cells $F$. Initial condition was taken to be $(A_1(0),A_2(0),F(0),T_{in}(0),T_1(0),T_2(0),V(0))=(0.9,0.0333,0,0.9,0,0,0.05)$ for all simulations, representing the fact that before the infection there are no infected cells and no activated T cells.

Figure~\ref{fig11} shows the dynamics of system (\ref{nondim}) for the case $p_1=p_2=1$, which corresponds to a situation when both types of cells $A_1$ and $A_2$ are targets of both infection and autoimmunity. For sufficiently small values of infection rate $\lambda$, the infection is being completely cleared: after the initial peak, the number of infected cells is monotonically decreasing, and the system approaches a stable steady state $E_8^*$, as shown in plots (a) and (b). This is the case of a normal disease clearance, where immune response of the host is able to completely clear the infection without causing an autoimmune reaction. Due to the exhaustion of the pool of infected cells, the activation of na\"ive T cells stops, and the population of activated T cells is then slowly diminishing. The same happens to autoreactive T cells, whose number reaches its peak slightly later than the population of activated T cells, as these cells have to be derived from the population of regular activated T cells $T_1$. It is worth noting that unlike an earlier model of Iwami {\it et al.} \cite{Iw1,Iw2}, the present model is able to support the initial viral expansion and subsequent clearance of infection by the immune system.

For higher values of $\lambda$, the disease-free state $E^{*}_8$ loses stability, and the system evolves to a stable equilibrium $E^{*}_{14}$, which describes the state of chronic infection. In this case, the immune system of the host is unable to clear the infection, and as a result it persists at a constant level. As $\lambda$ increases further, the steady state $E^{*}_{14}$ loses stability via Hopf bifurcation, giving rise to stable periodic oscillations, as shown in plots (e) and (f). In this case, one observes 
episodes of high viral production (relapses) with long periods of quiescence (remissions). Such dynamics have been observed in a number of autoimmune diseases, such as MS, autoimmune thyroid disease, uveitis etc. \cite{BF,D,Nyl}. An important note here is that none of the subsequent reactivations of the virus requires any exogenous factors, but rather the system itself cycles through periods of relative quiescence and viral release.
\begin{figure}
\hspace{-1cm}
\epsfig{file=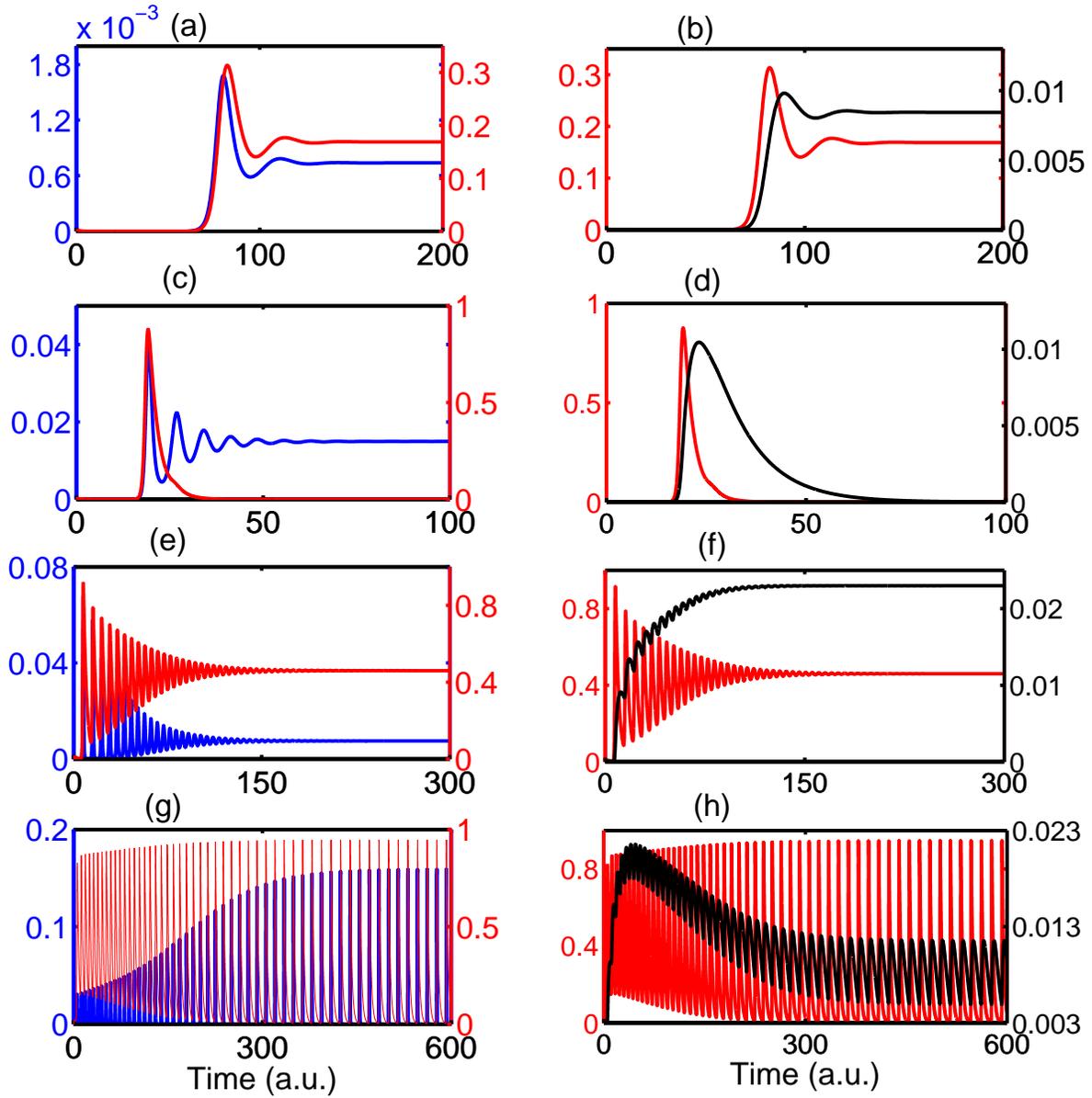,width=18cm}
\caption{Numerical solution of the system (\ref{nondim}). Parameter values are the same as in Fig.~\ref{fig11}, except for $p_1=p_2=0$. (a) and (b) $\lambda=0.9$, (c) and (d) $\lambda=1.5$, (e) and (f) $\lambda=5$, (g) and (h) $\lambda=9$. In all plots, the colours denote infected cells $F$ (blue), activated T cells $T_1$ (red), and autoreactive T cells $T_2$ (black). Time is measured in arbitrary units (a.u.).}\label{fig00}
\end{figure}
During each viral episode, the number of infected cells increases, and this triggers rapid activation of na\"ive T cells, which in turn suppresses viral production, leading to a decrease in the number of infected cells. As the autoreactive T cells in the simulation shown in Fig.~\ref{fig11} have a much longer lifetime, their number decreases much more slowly during the periods of relative suppression of infection. For even higher values of $\lambda$, this periodic solution becomes unstable, and instead the system evolves into another periodic orbit arising from the Hopf bifurcation of the steady state $E^{*}_{12}$, in which case the second population of target cells $A_2$ goes to zero.

Next, we consider another biologically plausible scenario when cells $A_2$ are only targeted by infection, and cells $A_1$ are only affected by autoimmunity, which is described by $p_1=p_2=0$. Figure~\ref{fig00} illustrates the dynamics of the system (\ref{nondim}) in this case. For sufficiently small values of $\lambda$, one observes normal clearance of infection similar to the case $p_1=p_2=1$, and as $\lambda$ increases, the system tends to a stable state of constant chronic infection $E^*_{14}$, as shown in plots (a) and (b). When the steady state $E^*_{14}$ loses stability, rather than develop sustained oscillations as in the case $p_1=p_2=1$, now the system goes instead to a stable steady state $E^*_{13}$ shown in plots (c) and (d), which has $T_{in}=T_1=T_2=0$. This situation describes a state, in which the numbers of na\"ive, activated and autoreactive T cells are all zero. While this is not biologically realistic, functionally it resembles exhaustion in which virus specific T cells are rendered ineffective and therefore the effective population size is reduced to zero. For higher values of $\lambda$, there is another stability switch, and the system evolves toward a stable equilibrium $E^*_{10}$. This behaviour is shown in plots (e) and (f), and it describes a situation when the first population of target cells $A_1$ goes to zero. When $\lambda$ is increased further still, the steady state $E^*_{10}$ becomes unstable, and one observes a state of autoimmunity represented by stable periodic solution arising from a Hopf bifurcation of a chronic steady state $E^*_{14}$.

In the case when only one of $p_1$ and $p_2$ is different from zero, the system can exhibit behaviours similar to the cases described above, with transitions between stable steady states and periodic solutions of different origins. Numerical simulations suggest that in most cases, for sufficiently large values of the infection rate $\lambda$, the system approaches a stable periodic orbit arising from the Hopf bifurcation of the steady state $E^*_{14}$. Such periodic solution corresponds to the above-mentioned state of autoimmunity with remissions and relapses.

\subsection{Multiple infections}

Since for many viral infections it is realistic to expect subsequent exposures of a person to the same virus, we now consider a scenario where someone who has recovered from a primary infection or currently has a chronic viral infection experiences a secondary viral challenge with the same virus. It is known that the timing of secondary infection plays an important role in determining the progress of infection, as well as the immune dynamics \cite{Nau,Sel}.

\begin{figure}
\hspace{-0.5cm}
\epsfig{file=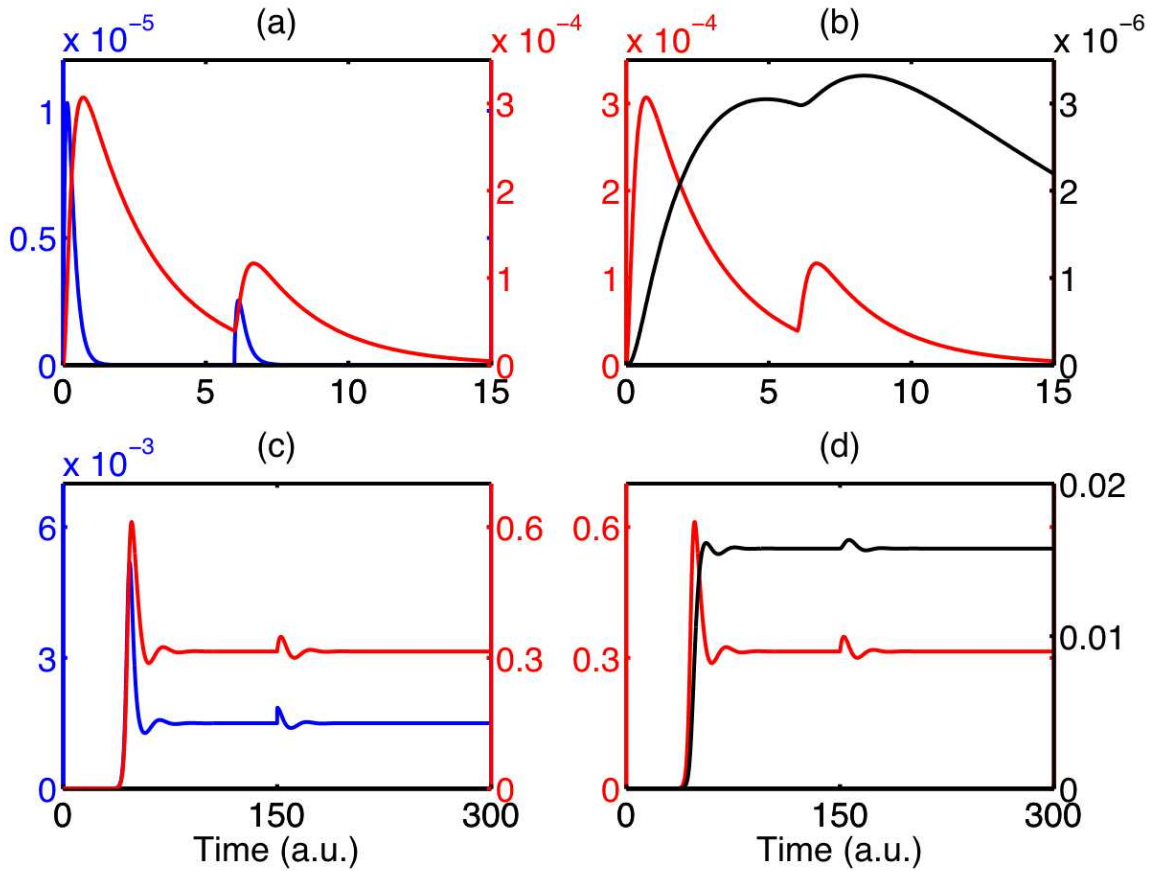,width=17cm}
\caption{Numerical solution of the system (\ref{nondim}). Parameter values are the same as in Fig.~\ref{fig11}, except for $p_1=p_2=0$. (a) and (b) $\lambda=0.2$, (c) and (d) $\lambda=1$. In all plots, the colours denote infected cells $F$ (blue), activated T cells $T_1$ (red), and autoreactive T cells $T_2$ (black). Time is measured in arbitrary units (a.u.).}\label{figmult}
\end{figure}

Figure ~\ref{figmult} illustrates the dynamics during a secondary viral exposure in the parameter regime with normal clearance of infection and during the chronic infection. In the case of normal clearance shown in plots (a) and (b), one can observe that due to a much slower decay of activated T cells, they still remain at a non-negligible level following the primary infection. This means that the second infection produces a significantly smaller number of infected cells. The later a secondary viral challenge occurs, the higher will be the resulting number of infected cells, and correspondingly the higher will be the numbers of activated and autoreactive T cells during a secondary infection. We find that the number of $T_2$ cells during a secondary infection is often higher than during a primary infection, such that the exacerbation of the autoimmune reaction is relatively much greater than the antiviral response, as shown in plot (b). The actual level of $T_1$ and $T_2$ cells depend on the timing of secondary infection: if a secondary infection takes place sufficiently close in time to the original infection, the number of $T_1$ cells will only increase by a very small amount, while the number of $T_2$ cells will exceed that number during the primary infection. If, however, a greater time elapses between the two infections, the number of activated T cells $T_1$ will be slightly greater, but still it will never exceed the number of $T_1$ cells during a primary episode. At the same time, the number of cross-reactive T cells $T_2$ in this case will be lower than during the primary infection.

Plots (c) and (d) show that when the system is chronically infected, due to the significant amount of activated T cells, the secondary infection does not lead to a major increase in the number of infected cells, and as a result the infection is quickly cleared to the same chronic level as before the secondary infection. We have also analysed the influence of secondary exposures on the dynamics of recurrent infections, and in this case the effect of subsequent viral challenges is quite small in that it does not change the amplitude or period of oscillations except producing a small additional peak in the number of infected cells immediately after the infection. From this we conclude that in the parameter regime when the infection is recurrent, the main role is played by the periodic nature of the system itself, and it is this that causes relapses and remissions, rather than subsequent infections.

\subsection{Treatment}

As a next step in the analysis of onset and dynamics of autoimmunity, it is instructive to consider a practically important issue of therapeutic intervention. Numerical simulations presented earlier suggest that in the case when autoimmunity is triggered by infection with a virus, there is no further need for infection to maintain the periodic state of relapses and remissions. Let us now consider how the autoimmune dynamics changes upon the introduction of a therapy, aimed at reducing the number of autoreactive T cells. It will be assumed that such treatment can have two potential impacts: it can either act by eliminating the autoreactive T cells, thereby reducing the overall burden of autoimmunity, or it can transform those autoreactive T cells into cells with a higher activation threshold, i.e. activated T cells. These scenarios correspond loosely to treatments with therapeutic monoclonal antibodies that delete populations of cells \cite{Coles}, or with drugs such as cyclosporin that inhibit the activation of T cells \cite{Liu}. In clinical studies, these therapies cannot yet be made to selectively target autoreactive cells, which remains an important ongoing goal of treatment. Mathematically, we model such an intervention by modifying the fifth and sixth equations of the system (\ref{model}) (and, correspondingly, (\ref{nondim})) as follows
\begin{equation}
\begin{array}{l}
\displaystyle{\frac{dT_{1}}{dt}=\alpha_{act}T_{in}F-T_{1}(\mu_{1}+\tau)+p_3CT_2\theta(t-t_0),}\\\\
\displaystyle{\frac{dT_{2}}{dt}=\tau T_{1}-\mu_{2}T_{2}-CT_2\theta(t-t_0),}
\end{array}
\end{equation}
\begin{figure}
\hspace{-1cm}
\epsfig{file=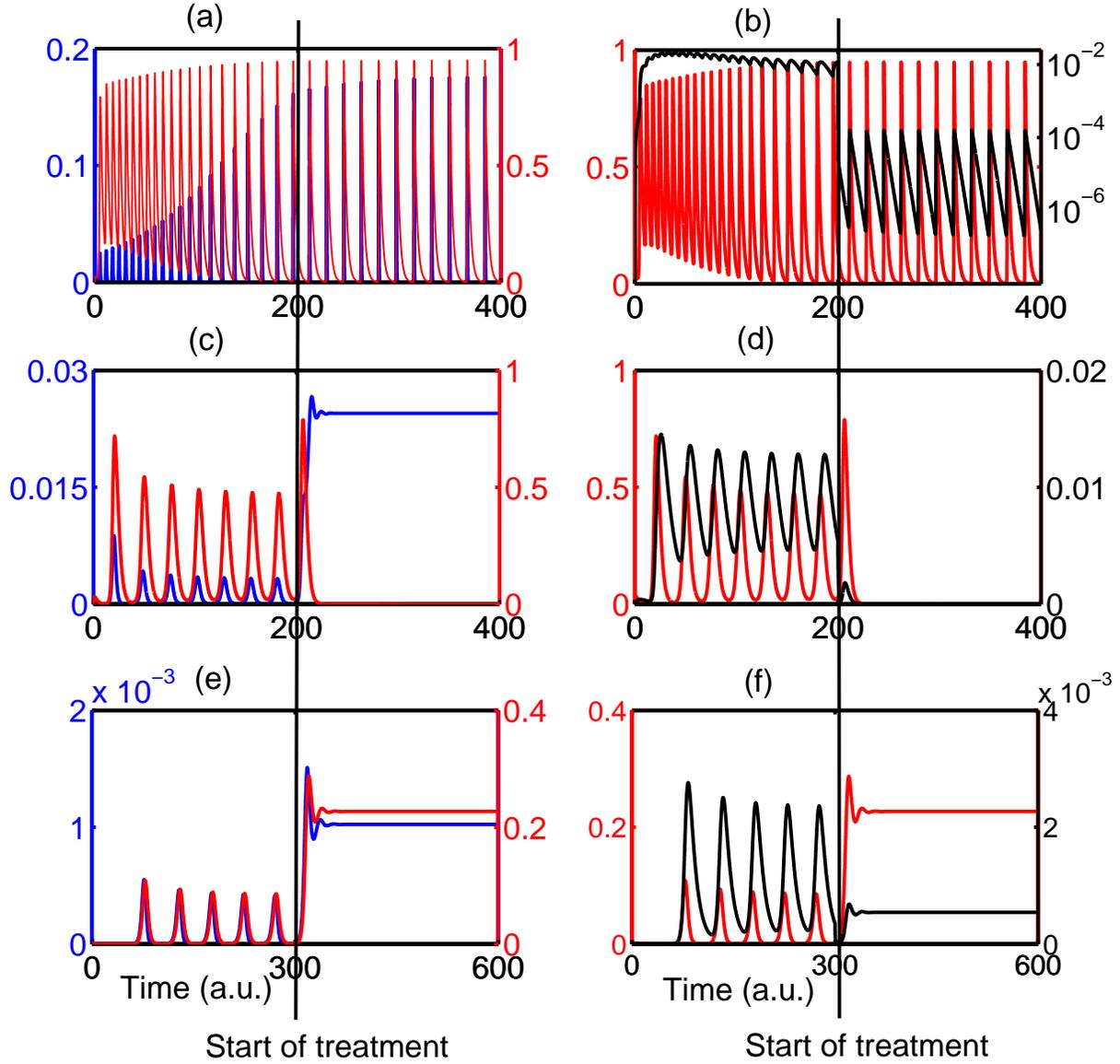,width=17.5cm}
\caption{Temporal dynamics during treatment of an autoimmune state. Parameter values are the same as in Fig.~\ref{fig11}, and $p_3=0$. (a) and (b) $p_1=p_2=0$, $\lambda=10$, $C=30$; (c) and (d) $p_1=1$, $p_2=0$, $\lambda=0.65$, $C=2$; (e) and (f) $p_1=0$, $p_2=1$, $\lambda=0.9$, $C=2$. Time is measured in arbitrary units (a.u.).}\label{tper}
\end{figure}
\noindent where $C$ is the rate at which autoreactive $T_2$ cells are destroyed by treatment, $p_3$ shows the fraction of autoreactive T cells that can be converted into activated T cells (when $p_3=0$, the treatment is only responsible for reducing the number of $T_2$ cells), $\theta(\cdot)$ is the Heaviside function, and $t_0$ is the time when the treatment is introduced. We assume that initially a person is exposed to a viral infection and develops some sort of autoimmune response, and in response to this a treatment is introduced in order to reduce or eliminate autoimmunity.

Figure~\ref{tper} shows the effects of treatment on the dynamics of immune response in the regime, where before treatment the system exhibited sustained periodic oscillations corresponding to relapses and remission during autoimmunity. All the simulations were performed for the case $p_3=0$, but it has been checked that the same results are obtained for $p_3>0$. In the case when $p_1=p_2=0$ (i.e. cells $A_2$ are the only target of infection, and cells $A_1$ are only affected by autoimmunity), once the treatment is introduced, the number of autoreactive T cells reduces significantly, as shown in plots (a) and (b). Although the oscillations persist after treatment, the number of $A_1$ is greatly increased, and for sufficiently high rate of treatment $C$, it stays very close to 1. For $p_1=1$ and $p_2=0$ (infection targets both $A_1$ and $A_2$ cells, but only $A_1$ cells are affected by autoimmune response), plots (c) and (d) indicate that the introduction of treatment leads to a stable steady state $E^*_{13}$, in which all T cells ($T_{in}$, $T_1$ and $T_2$) are eliminated. It is noteworthy that whilst the autoimmune reaction is eliminated in this scenario, this also leads to an increased level of persistent infection. When $p_1=0$ and $p_2=1$ (only $A_2$ cells are a target of infection, and both $A_1$ and $A_2$ cells are affected by autoimmunity), treatment leads to suppression of oscillations and establishment of a stable chronic state $E^*_{14}$, as illustrated in plots (e) and (f). Due to a high rate $\alpha_a$, at which autoreactive T cells $T_2$ destroy infected cells, once the population of these cells is reduced, the resulting chronic state $E_{14}$ s characterized by a higher level of infected cells. In the case when $p_1=p_2=1$, behaviour of the system under treatment is qualitatively similar to the case of $p_1=1$ and $p_2=0$, when all T cells are eliminated.

\begin{figure}
\epsfig{file=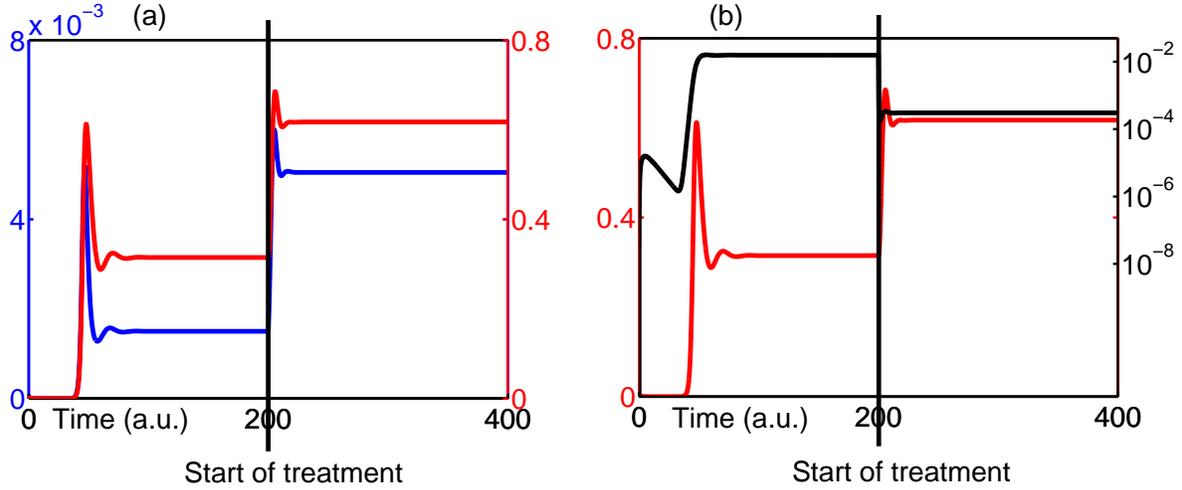,width=16cm}
\caption{Temporal dynamics during treatment of a chronic infection. Parameter values are the same as in Fig.~\ref{fig11}, $p_1=p_2=p_3=0$, $\lambda=1$, and $C=30$. Time is measured in arbitrary units (a.u.).}\label{tchron}
\end{figure}

In Fig.~\ref{tchron} we illustrate how treatment affects the dynamics in the case of chronic infection. One can observe that treatment leads to a significant reduction in the number of autoreactive T cells, prompting a substantial increase in the number of $A_1$ cells. However, treatment does not completely eliminate the infection which remains chronic, and similarly to the treatment of an autoimmune state, while the treatment reduces the level of autoimmune reaction, it simultaneously leads to a relative increase in the number of infected cells. This is biologically reasonable since inhibiting autoimmunity can lead to the reactivation of infection \cite{Hell}.

\section{Discussion}

In this paper we have developed and analysed a mathematical model for the dynamics of the immune response against a viral infection and the associated onset of autoimmunity. Having introduced separate populations of target cells that can be affected by infection and/or autoimmunity, as well as different compartments for T cells with different activation thresholds, we have studied how the outcome of the immune response can be the clearance of infection, chronic infection or recurrent infection. In the case of normal clearance, there are no lasting immune consequences for the organism. The number of autoimmune T cells is small and decreases with time after the clearance of infection. Chronic infections are characterized by a constant level of activated and autoreactive immune cells, which keep infection in check but do not clear it. Another possible scenario for persistent infection is when virus specific T cells is exhausted, and hence the effective T cell population is reduced to zero. Recurrent infections are characterized by inability of the immune response to clear an infection, which results in relatively long periods of remission followed by relapses with a substantial viral production and a large number of cells being infected.  In this study we have focussed on chronic infection driven by an increase in the infection rate $\lambda$. Studies in other models show that chronicity can also be influenced by other factors such as the frequency of antigen specific T cells \cite{NN}. During recurrent infection we observe higher numbers of autoimmune cells (compared to the case of chronic infection), which exacerbate the relapses and cause more damage. Notably, there is a difference in the outcome of the infection, depending on whether infection and autoimmunity affect the same or different organs.

Numerical solutions reveal a number of potentially important features in the dynamics of the autoreactive cells. When a person experiences a secondary challenge with the same virus, the model shows that in the case of normal clearance the number of cells getting infected is smaller in subsequent infections due to a limited number of T cells remaining activated at the time of secondary infection. Although in this case the infection is still successfully cleared, 
in a number of circumstances the second and subsequent peaks in autoreactive cell numbers exceeded that seen during the initial response to virus (Fig.~\ref{figmult} (b)), thus causing an additional immune challenge. This could implicate repeated cycles of T cell expansion, or multiple linked rounds of infection as crucially underpinning the development of frank autoimmune disease. This would be consistent with multi-hit models of autoimmunity (e.g. \cite{Amb}) and also with circumstances in which more than one infectious episode is necessary to precipitate frank disease. In contrast, secondary infection during a chronic infection does not have a significant effect on the dynamics. The same conclusion holds for the recurrent infections, thus indicating that in this case it is the intrinsic dynamics of the interactions between the immune system and the virus that causes remissions and relapses rather than the fact that a person experiences further infections.

In the form of the model where the population of infected cells is different from that which is the target of autoimmunity, we found that under some conditions, low levels of viral persistence could be associated with high levels of autoreactivity (Fig.~\ref{fig00} (e)\&(f)). We also demonstrated that the system can approach a state that resembles T cell exhaustion (Fig.~\ref{fig00} (c)\&(d)). This is intriguing because inactivation of the immune response in the face of chronic viral infection and in tumours is a well described and important area of ongoing investigation \cite{Kim}.

We have also studied the effects of treatment aimed at reducing the number of autoreactive cells on the dynamics of the immune response. In the case of the recurrent autoimmune state, when infection and autoimmunity occur in different organs, such treatment leads to a substantial improvement of the situation, significantly reducing the number of autoreactive T cells. It is important to note that although this does not eliminate episodes of relapses/remissions, they have a much less prominent impact on the numbers of susceptible cells and hence cause significantly less damage than the full-blown autoimmunity. When only one of the susceptible cell populations is the target of infection but both of them can be affected by autoimmunity, introduction of treatment leads to suppression of relapse/remission oscillations and establishment of a state of chronic infection. When initially the infection is chronic, treatment does not qualitatively change the dynamic state of the system, but leads to a reduction in the number of autoreactive T cells. One should note, however, that due to reliance of immune system on autoreactive T cells to contribute to control of infection, any treatment aimed solely at reducing the number of these cells can inadvertently lead to a higher level of persistent chronic infection. This trade-off between the ability to control the infection and at the same time to minimize the undesired effects of autoimmunity is somewhat similar to the problem in chemotherapy where an effective treatment of a tumor may have a negative impact on the overall immune status, thus requiring some sort of adjuvant therapy. At the same time, if the role played by autoreactive T cells in the clearance of infection is not so significant, the above problem becomes less serious, and it is possible to achieve efficient control of autoimmunity without compromising host's ability to fight infection.

There are a number of interesting potential extensions to this work which may be possible in the future. While we started with the concept of tunable activation thresholds, the implementation we chose here was deliberately simplified to ensure the tractability of the model. Previous theoretical and experimental work \cite{AG,AW,GP,GS,GP2,VdB04} has stressed the dynamic nature of the tuning process which can be observed over timescales of minutes to days \cite{Rom,SDG02}. It would therefore be interesting to explore the dynamics of this process and the effects that different parameters for the kinetics of tuning might have on the development of autoreactive T cell populations. Another feature of the current model is that the autoimmune response does not persist when the virus is cleared. Since there is good experimental evidence that autoimmune responses can be self-sustaining and chronic \cite{Kerr}, developing the model to explore this behaviour will be an important further development of the model.

\section*{Acknowledgements}

We would like to thank two anonymous referees for their comments and suggestions, which have helped to improve the presentation in this paper.

\end{document}